\documentclass[lettersize,journal]{IEEEtran}

\ifCLASSINFOpdf
  \usepackage[pdftex]{graphicx}

\else
  
\fi

\usepackage{amsmath}
\usepackage{array}
\usepackage{booktabs,tabularx}

\usepackage{algorithm}
\usepackage{algpseudocode}

\usepackage{stfloats}
\usepackage{graphicx}
\usepackage{amsmath}
\usepackage{resizegather}

\usepackage{url}

\usepackage{float}
\usepackage[many]{tcolorbox}
\usepackage{xcolor}

\usepackage[%
    colorlinks=true,
    pdfborder={0 0 0},
    linkcolor=black
]{hyperref}

\hypersetup{citecolor=black}

\begin{document}

\title{LiPI: Lightweight Privacy-Preserving \\Data Aggregation in IoT}

\author
{\IEEEauthorblockN
{Himanshu Goyal,
Krishna Kodali,
Sudipta Saha\\
\IEEEauthorblockA{School of Electrical Sciences,
Indian Institute of Technology Bhubaneswar, India\\ 
Email: \{hg11, knk11, sudipta\}@iitbbs.ac.in}}
}

\maketitle

\begin{abstract}
\label{sec:abstract}
In the modern digital world, a user of a smart system remains surrounded with as well as observed by a number of tiny IoT devices round the clock almost everywhere. Unfortunately, the ability of these devices to sense and share various physical parameters, although play a key role in these smart systems but also causes the threat of breach of the privacy of the users. Existing solutions for privacy-preserving computation for decentralized systems either use too complex cryptographic techniques or exploit an extremely high degree of message passing and hence, are not suitable for the resource-constrained IoT devices that constitute a significant fraction of a smart system. In this work, we propose a novel lightweight strategy \textit{LiPI} for \textit{Privacy-Preserving Data Aggregation} in low-power IoT systems. The design of the strategy is based on decentralized and collaborative data obfuscation and does not exploit any dependency on any \textit{trusted third party}. In addition, besides minimizing the communication requirements, we make appropriate use of the recent advances in  \textit{Synchronous-Transmission} (ST)-based protocols in our design to accomplish the goal efficiently. Extensive evaluation based on comprehensive experiments in both simulation platforms and publicly available WSN/IoT testbeds demonstrates that our strategy works up to at least 51.7\% faster and consumes 50.5\% lesser energy compared to the existing state-of-the-art strategies.  
\end{abstract}

\section{Introduction}
\label{sec:intro}

In today's modern digital world, many vital and significant aspects of human lives are being substantially driven by IoT/WSN-assisted smart systems, e.g., Smart-Health-Care \cite{7508172}, Wireless-Body-Area Networks \cite{7950903}, Intelligent-Transportation \cite{7063743, 8037972}, Smart-Grid \cite{8701687}, and many others. 
Due to the close interaction of the smart systems with human life, the data sensed by these devices often bear a direct relationship with the private/sensitive information of the users. For instance, in \textit{Advanced Metering Infrastructure} (AMI) \cite{7841782}, the electricity consumption data collected from a house can be used to precisely infer the activities inside the house \cite{desai2019survey}. Similarly, leakage of the values of the raw physical parameters sensed by the IoT devices used in the deployment of \textit{the Wireless Body Area Network} (WBAN) can also be used to infer the various vital health-related status of a patient. 
IoT applications, therefore, instead of 
using the raw data, adopt various \textit{privacy-preserving} strategies where the data are appropriately changed before being shared by the devices, and take necessary steps to make use of it to accomplish the final goal. 

In this work, we address the \textit{Privacy-Preserving Data-Aggregation} (PPDA) strategies. Data aggregation is a very simple, common, and frequently executed operation in any smart system. Ideally, the data are first shared by the devices, and then the target aggregation function is applied to get the result. However, privacy preservation strategies make it complicated by incorporating additional steps to change the data and also arranging necessary means to make use of the same. Since a significant fraction of the devices used in a smart system are resource-constrained, the use of complex strategies for privacy preservation either makes it impossible or at least harder for a task to get completed fast which ultimately hampers the behavior of a smart system. Thus, the design of an efficient PPDA strategy for resource-constrained devices is a challenging task.

A number of works in PPDA are based on \textbf{Homomorphic Encryption} \textbf{(HE)} \cite{gentry2009fully} which enables computation of aggregation operations directly over cipher-text alleviating the requirement of the intermediate deciphering of the data. However, in many HE-based works \cite{LPDA, LVPDA, PEPPDA, 3PDA, PEDA}, the sink (i.e., the final destination of the data) is assumed to be trustworthy and is provided with the key to decipher the final result, which also enables it to decipher the individual cipher-texts from the individual nodes too. To resolve this, the work 
\textbf{(PEPPDA)} \cite{PEPPDA} uses a tree structure and dynamic slicing of the data. The work  
\textbf{(3PDA)} \cite{3PDA}, in contrast, uses an intermediate \textit{Data Collection Unit} to achieve the same. However, apart from these known issues, in general, execution of HE is quite computationally intensive which is hence not suitable for low-power IoT systems.

\textbf{Multi-Party Computation (MPC)} has been adopted for solving PPDA by a set of works. The concept of \textit{Secure Multi-Party Computation} (SMPC) was introduced by the work \cite{4568388}. \textit{Shamir's Secret Sharing} \cite{10.1145/359168.359176} is one of the most adopted strategies to achieve SMPC in decentralized systems. In SMPC, \textit{first}, each source node divides its secret value into multiple \textit{parts} and shares the different parts with different nodes. Finally, the node runs the second round of data-sharing to derive the final result. The algorithms 
\textbf{CPDA} and 
\textbf{SMART} proposed in work \cite{4215819} improve upon these basic concepts through the application of \textit{secret-splitting} and collaborative computing. MPC-based approaches, thus, in general, exploit less computation but incurs heavy communication over secure channels, which is also a serious concern in the context of energy-constrained IoT systems.

\newpage
Appropriate obfuscation of the data (referred to as \textbf{Data Obfuscation} or \textbf{DO}) with the help of random noise has also been a very standard way of hiding the original data to achieve PPDA \cite{HFAPPA, 10.1007/11761679_29, DPPDA}. However, the noise is supposed to be generated in a very organized way so that it can be effectively removed to obtain the correct aggregation. Most of these DO-based works, hence depend on some \textit{Trusted Third Party} (TTP) for sharing the necessary secret keys and random noise values which makes them unreliable. Some of the works, e.g., \textbf{DPPDA} \cite{DPPDA}, generate the noise values without the help of any TTP and achieves accurate convergence. However, it is applicable only for finding \textit{average} and it also exploits the knowledge of the network topology which may not be available in a generic setting.

In this work, we propose a novel decentralized lightweight PPDA strategy, \textit{LiPI}, which fundamentally makes use of \textit{data obfuscation through decentralized collaboration among the devices}. LiPI is designed as a self-sufficient decentralized strategy that does not depend on any TTP or any centralized entity for any purpose. 
It also does not use any heavyweight cryptographic mechanism or explicit pair-wise secret channel to achieve PPDA. In addition, in this work, for the first time, we employ a \textit{Synchronous-Transmission} (ST) based framework for efficiently handling the communication requirements to achieve PPDA in resource-constrained IoT devices.
Moreover, most of the works on PPDA carried out so far are either theoretical or simulation-based. In contrast, we implement our protocol in real low-power off-the-shelf IoT devices and also evaluate it over publicly available IoT/WSN testbeds. 

The main contributions of the proposed work are summarized below. 

\begin{itemize}
    
    \item We design a simple lightweight strategy LiPI for PPDA in decentralized resource-constrained IoT systems with unsecured communication links based on collaborative data-obfuscation.
    
    \item We propose an ST-based framework to efficiently meet the communication requirements to accomplish PPDA in resource-constrained IoT systems.
    
    \item We implement LiPI as well as a few other state-of-the-art strategies for PPDA in our proposed framework with the help of Contiki OS for TelosB devices and rigorously test them through simulation and publicly available WSN/IoT Testbeds.
    
\end{itemize}

The rest of the paper is organized as follows. Section~\ref{sec:related} provides a concise description of the pros and cons of various prior works. Section~\ref{sec:adversary-model} describes the adversary model and the basic goals of the current work. Section~\ref{sec:background} provides a brief description of the ST-based protocols that are used to construct the proposed system framework of LiPI. Section~\ref{sec:design} provides a theoretical description of the proposed strategy along with its algorithmic design. Section~\ref{sec:failure} talks about the design of the system considering the possibility of node failures and its relevant mitigation techniques. Finally, Section~\ref{sec:evaluation} reports the evaluation of LiPI in comparison with the other existing strategies.

\begin{table*}
    \vspace*{-0.7cm}
     \caption{Summary of the existing approaches achieving Privacy-Preserving Data Aggregation.}
    \includegraphics[width=\textwidth]{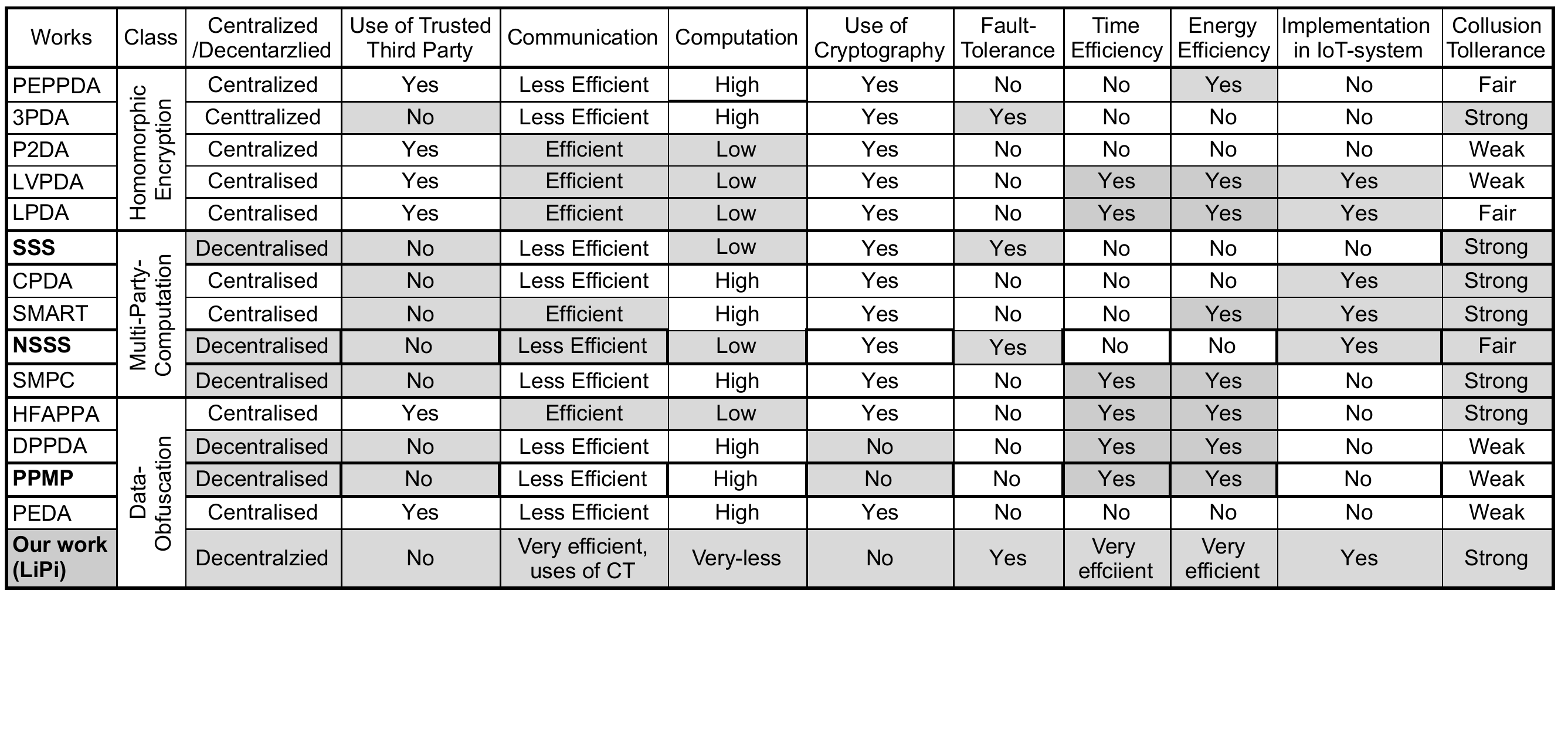}
    \label{fig:workscomparison}
    \vspace{-1.5cm}
\end{table*}

\section{Related works}
\label{sec:related}

In the following, we review the existing works on PPDA \cite{abbasian2020survey,POURGHEBLEH201723,OZDEMIR20092022} based on their adaptability in a resource-constrained IoT system.

\textbf{Trusted Third Party (TTP):} Use of secure communication is quite common in PPDA. In most cases, a TTP is assumed for the necessary key distribution \cite{LPDA, LVPDA, PEDA, P2DA}. However, such dependency opens up a channel for leakage of private information through the TTP itself.  
In the current work, we demonstrate an efficient strategy to generate the keys on-the-fly through a collaborative mechanism that absolutely avoids the use of any TTP.

\textbf{Centralized/Decentralized:} Most of the HE-based approaches use cloud-based/centralized service for heavy computation tasks (e.g., use of \textit{Base Station} (BS)) \cite{PEPPDA, 3PDA, PEDA}. SMPC-based works \cite{8406322} generally use a decentralized model, but they rely on a secure channel for communication. The work \textbf{PPMP} \cite{PPMP} uses a decentralized setting along with an unsecured channel. It follows a collaborative data-obfuscation strategy as LiPI. Section~\ref{subsec:compareimplementation} describes the operation of PPMP in brief. We implement PPMP in our framework and compare its performance with LiPI.

\textbf{Cryptography:} Most of the PPDA works that are targeted to IoT systems try to minimize the use of cryptography (except a few, e.g., \cite{POURGHEBLEH201723, OZDEMIR20092022, iccsss,goyal2022multiparty}) as they are not suitable for resource-constrained devices. Lightweight cryptography is also used in various works, e.g., 
\textbf{LPDA} \cite{LPDA} and 
\textbf{LVPDA} \cite{LVPDA}. However, these approaches are centralized and rely on a TTP. In the current work, we do not use any cryptography and assume unsecured channels. The work PPMP \cite{PPMP} also considers a purely unsecured environment. However, the design of PPMP involves more communication and also lack several other features which we discuss in detail in Section~\ref{sec:evaluation}.

\textbf{Collaborative-computation:} Dependency over centralized computation is quite common in PPDA strategies. However, it incurs common problems like single-point-of-failure and also limits scalability. Collaborative computation, as used in the algorithms CPDA and SMART proposed in work, \cite{4215819} completely avoids such problems. However, such works still partially rely on a centralized entity like BS for computing the final aggregated value. The work \cite{6208686} realizes SMPC in a purely decentralized environment. However, it relaxes the original problem by assuming that the data can be shared through a \textit{secured channel} between any pair of nodes within the network which is not realistic in low-power wireless systems. Collaborative computation has been adopted by DO-based works such as  
\textbf{HFAPPA} \cite{HFAPPA}. However, they mostly depend on a TTP for system initialization. In this work, we design a self-sufficient collaborative computation in a purely decentralized setting and without using any secured channels.

\textbf{Fault-tolerance:} Fault-tolerance is a very significant issue that is rarely addressed in the existing PPDA works. Sudden drops of multiple nodes are quite common in IoT systems and may result in wrong computation of the result in even rigid strategies like 
PPMP \cite{PPMP}. The work \cite{chen2015pdaft} considers fault tolerance in Smart-Grids. However, it relies on TTP in case a node drops. In contrast, our design not only automatically detects any node failure but also takes appropriate action to resolve it.

\textbf{Communication:} The works based on MPC and DO \cite{3PDA, 4215819, HFAPPA, DPPDA, 6208686}, exploit heavy communication among the nodes causing high message complexity. The recent work \textbf{SSS} \cite{iccsss} in this direction applies the concept of Shamir's Secret Sharing to achieve PPDA in a practical setting. However, due to the high communication cost, it does not fit with an IoT setting. The work \textbf{NSSS} \cite{goyal2022multiparty} attempts to improve the communication cost of SSS and makes it suitable for IoT up to a certain extent. Our design optimizes the communication requirements. In addition, it exploits the recent advancements in ST-based strategies to efficiently achieve the goals. We discuss both SSS and NSSS in detail in Section ~\ref{subsec:compareimplementation} and re-implement them in our framework to compare their performance with LiPI.

Finally, the majority of the existing works either demonstrate simulation or theoretical validation of their concepts or test their proposed solutions on resource-rich systems. 
In contrast, in the current work, we demonstrate our proposed design in real low-power resource-constrained IoT devices and test it in publicly available IoT testbeds.

Table~\ref{fig:workscomparison} summarizes the pros and cons of the major works in the domain. In general, most of the works, although strong in some set of aspects, seriously lacking in some others. The gray color background of a cell indicates a specific aspect of the work that is in favor of a low-power IoT system. Our target in the current work is to address each of the issues necessary to develop a self-sufficient solution for PPDA in a low-power IoT system.

\begin{figure*}[htbp]
  \includegraphics[width=\linewidth]{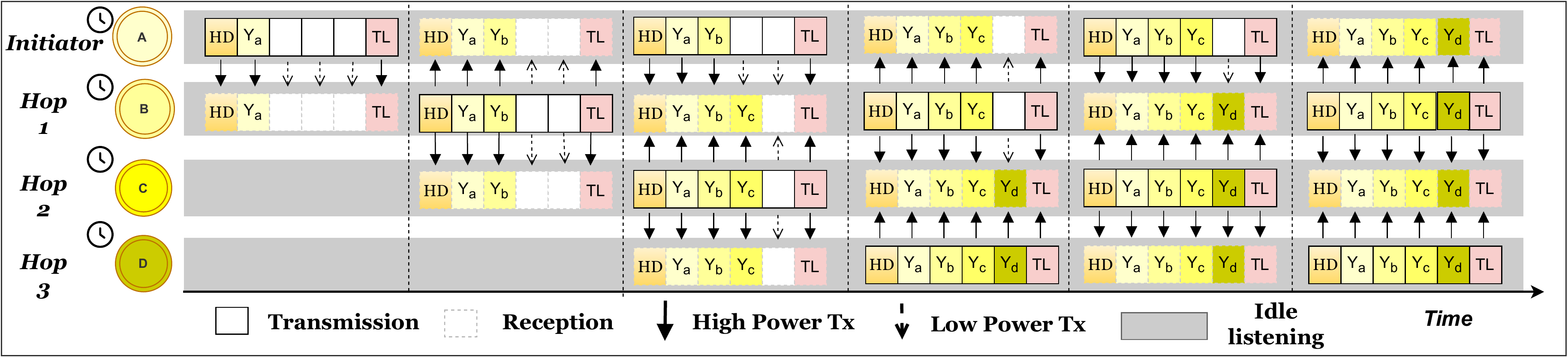}    \caption{Execution of all-to-all data-sharing using MiniCast in a 3-hop network.}
    \label{fig:minicast}
\end{figure*}

\section{Goals and the Basic Idea}
\label{sec:adversary-model}

We assume a \textit{semi-honest adversarial} model where every node in the system is supposed to follow the protocol specification correctly. However, the \textit{honest-but-curious} or \textit{passive} adversaries are free to learn the information from the internal states of the other nodes. 

The proposed PPDA strategy tries to fulfill the following goals. 

\textbf{Privacy:} Preservation of privacy of individual values shared by the nodes in the network even when any set of less than \(n-1\) parties collude ($n$ being the total no of participants), i.e., no one can learn the input of any other node in the network.

\textbf{Correctness:} If all the participating nodes are semi-honest, then the overall aggregated value is the target joint function of these private values.
    
\textbf{Robustness:} The aggregated value remains intact even if some of the nodes drop out intentionally from the protocol after the initial bootstrapping phase.
    
\textbf{Simplicity:} Finally, we aim to make it compatible with tiny low-power IoT-edge devices and hence keep the algorithm and the protocol as less computationally intensive as possible by avoiding cryptography, heavy computation, etc.

To achieve these goals, we approach with lightweight and collaborative data obfuscation. In a nutshell, we make every node share its secret data after properly obfuscating it through a collaborative masking function and then share the masked values in plain text with each other. Next, the local application of a de-masking function over the set of obfuscated data obtained from all the nodes produces the target joint function value. Collaborative obfuscation is achieved through the generation of pair-wise noise values for which we use pair-wise secret keys. Note that these keys are not used for any specific encryption or decryption purpose. Rather they are generated on the fly and discarded after use or renewed when necessary. An efficient way is demonstrated to accomplish this also. Finally, to meet the communication/data-sharing requirements efficiently, we adopt an ST-based framework. In the following, we provide a brief description of ST-based protocols used in our work.

\section{Synchronous-Transmission(ST) based Data-Sharing}
\label{sec:background}

Sharing of data among each other is the key component in any collaborative/decentralized algorithm. Existing DO or MPC-based works mostly use traditional \textit{Asynchronous-Transmission} (AT) based strategies to serve the purpose. However, due to mostly uncoordinated and unplanned transmissions in AT, with even a little rise in the data traffic, the chance of packet collision rapidly shoots up, causing drastic degradation in the performance of the protocol. Such issues restrict the use of AT in cases when faster interaction among the devices is necessary, especially in multi-hop wireless networks where the chance of collision among packets is inherently high.

In many recent works, ST-based strategies have been shown to be superior to AT-based ones in terms of reliability, latency as well as energy consumption \cite{sync_survey}. In contrast to AT, the protocols under CT try to coordinate the transmissions from the nodes in such a way that the packets originating from different source nodes, instead of colliding with each other, successfully exploit certain physical layer phenomena, e.g., \textit{Capture-Effect} (CE)/\textit{Constructive-Interference} (CI) and get correctly received at the destination nodes. Due to substantial avoidance of collisions, ST-based strategies save a lot of time and energy in the devices. ST has been used in various all-to-all/many-to-many data-sharing protocols in IoT/WSN \cite{lwb,chaos,codecast,mixer, minicast}. Because of its simplicity and efficiency, in this work, we use the protocol MiniCast \cite{minicast} as the base of LiPI. The many-to-many data sharing protocols are mostly founded on another protocol, Glossy \cite{glossy}. For completeness, below, we provide a brief description of Glossy and the next MiniCast.

\textbf{Glossy}: Glossy \cite{glossy}, for the first time, demonstrates how ST can be achieved through software-based lightweight time synchronization and can be used for network-wide flooding in resource-constrained IoT/WSN systems.
Glossy starts with a designated initiator node broadcasting the first packet. It triggers the transmissions from the first-hop neighbors of the initiator. Unlike AT, the time-aligned transmissions from these nodes result in CE and get received by the second-hop nodes which in turn respond together. The process goes on this way until the full network gets covered.

\textbf{MiniCast}: 
MiniCast extends Glossy by enabling efficient sharing of data from all or many source nodes in the network. Fundamentally, it arranges an interspersed execution of multiple Glossy floods in a very compact and systematic way with the help of a TDMA schedule. The data packets are transmitted from each node in the form of a chain as per the schedule. The duration of the transmission/reception of a full chain of packets is referred to as a \textit{slot} where every node is given a unique position, i.e., \textit{sub-slot}. Fig.~\ref{fig:minicast} pictorially shows the execution of MiniCast in a three-hop network with the status of one representative source node from each hop in consecutive time steps. Including the initiator, total four nodes thus are willing to share their data items \(Y_{a}, Y_{b}, Y_{c}\) and \(Y_{d}\), with each other. HD and TL represent the header and trailer packets in the chain. They are used for marking the start and end of a chain. In summary, MiniCast starts with every source node $N_i$ willing to share data item $d_i$ and ends up with each node obtaining a vector $<d_0, d_1, ..., d_i,...d_n>$ ($n$ being the no of the source nodes).

\section{Algorithmic Design}
\label{sec:design}

Algo.~\ref{algo:controller} provides an overall sketch of LiPI. Let $S_i$ be the private data of node $N_i$. The target is to compute a function $f(S_1, S_2,...S_n) = S$ in each node without allowing a node to know the $S_i$'s from the other nodes. Every node first obfuscates $S_i$ using a function $f_1$ (referred to as \textit{masking} function) and obtains $M_{i}$. In other words, $M_i = f_1(S_i, q_{i1},...,q_{ij},...,q_{in})$, where $q_{ij}$'s are \textit{noise values} which are added by node $N_i$ for every other source node $N_j$ in the system. In the subsequent MiniCast round, each node shares $M_{i}$ and on completion, receives a vector $<M_1, M_2, ..., M_i, ..., M_n>$. Finally, the function $f_2$, referred to as the \textit{de-masking} function, is applied over this vector to produce the final outcome, i.e., the target joint aggregation function ($f$). Conversely, $S = f_2(M_1, M_2, ..., M_i, ..., M_n) = f(S_1, S_2, ..., S_i, ..., S_n)$. Part-2 (Aggregation) of Algo. \ref{algo:controller} describes this procedure.

\begin{figure*} %
    \vspace*{-1cm}
    \includegraphics[width=\linewidth]{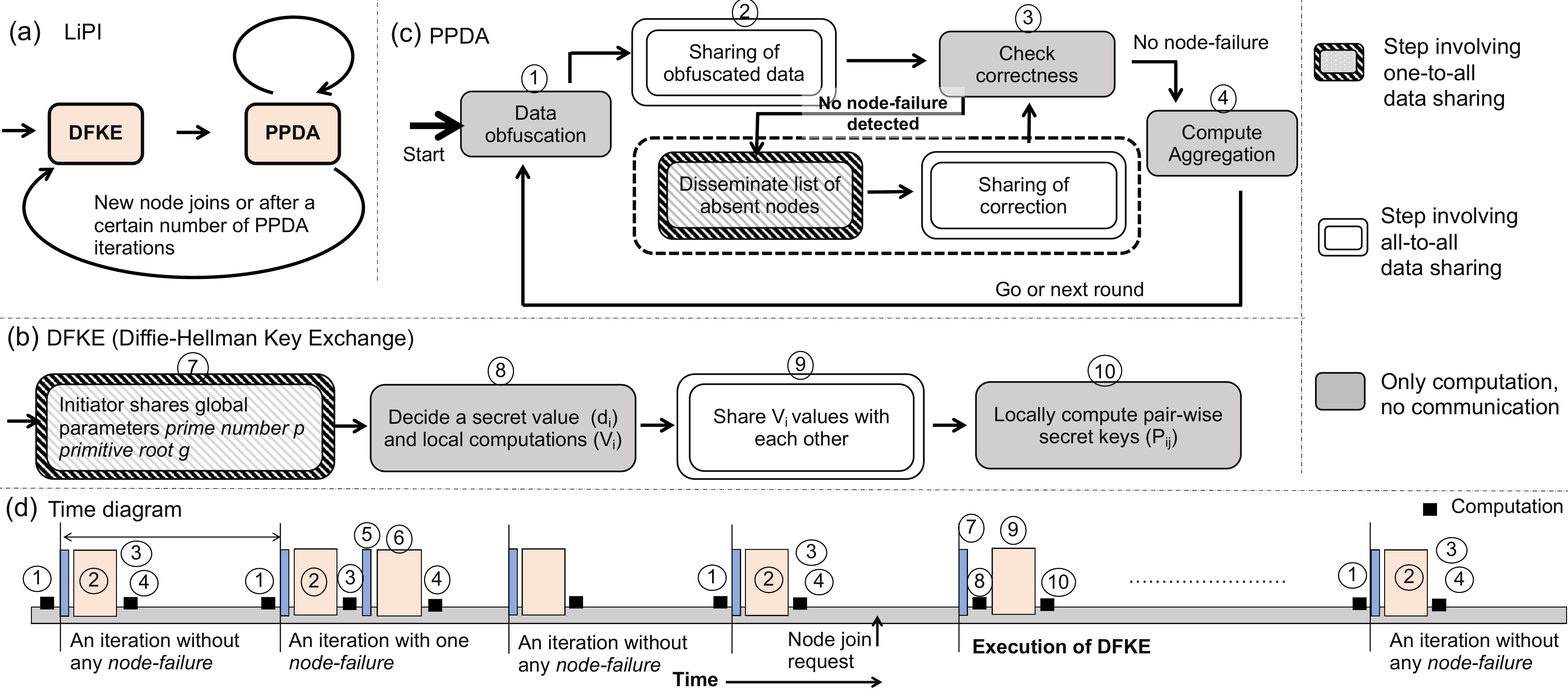}
    \caption{Schematic View of Proposed Methodology}
    \label{fig:state}
\end{figure*}

\textbf{Generating noise values:} A node computes a noise-value for every other node, e.g., the value $r_{ij}$ is calculated by node $N_i$ for node $N_j$. Pre-decided pair-wise secret keys (referred to as $P_{ij}$) are used for this purpose. We randomize the values of $r_{ij}$ over different iterations of PPDA using a \textit{Pseudo Random Number Generator} (PRNG) function where the MiniCast iteration number (i.e., seq\_no) concatenated with $P_{ij}$ is used as a seed, i.e., $r_{ij}$ is derived as PRNG($P_{ij} ||$ seq\_no) \footnote{MiniCast, like any other ST based protocol, runs in a periodic fashion where a sequence number (seq\_no) indicates how many data-sharing cycles are completed.}. 

\textbf{Collaborative-obfuscation:} Every node independently computes a noise value for every other node in the system as described above. The function $f_1$ obtains a single value $M_i$ combining the noise and its own private data, which is shared through MiniCast with other nodes. The noise quantities are computed in such a way that when all these $M_i$'s are collected together, the application of $f_2$ locally in each node can cancel each other, leaving only the $S_i$'s and hence correctly compute the aggregation. For example, if node $N_i$ contributes noise $q_{ij}$ for node $N_j$ and node $N_j$ contributes $q_{ji}$ for node $N_i$, then during the computation of $f_2$, $q_{ji}$ is supposed to cancel or nullify $q_{ij}$. To meet this, the noise quantities are specially modified through a third function $f_3$, i.e., $q_{ij}=f_3(r_{ij}$). Part-1 (Pre-processing) of Algo.~\ref{algo:controller} describes this procedure. Exact form of $f_3$ depends on both $f_1$ and $f_2$ as described with an example later.

\subsection{Pair-wise secret-key}

The obfuscated private values from each node are shared in plain text through MiniCast. A passive adversary $N_k$ can reveal the original secret value shared by $N_j$ if and only if it can re-generate all the pair-wise noise values used by $N_j$ that were used to obfuscate $S_j$. Use of pairwise secret keys while generating the noise values makes this hard or impossible. To avoid the use of any TTP, we generate the pairwise keys in a collaborative way by exploiting the well-known \textit{Diffie-Hellman Key Exchange} (DFKE) \cite{dfke} mechanism along with ST-based MiniCast. Algo.~\ref{algo:dfkey} narrates the key-exchange process. In a nutshell, the initiator globally decides the necessary parameters ($g$ and prime number $p$) and shares it with all the nodes using an instance of Glossy. Each node decides a secret value ($d$) locally and computes $V_i = g^{d_i}$ $mod$ $p$ which are shared among each other through a round of MiniCast. Finally, these values are used as per the standard DFKE process to compute the pair-wise secret keys. At the end of DFKE, $N-1$ pairwise secret keys are set for each node. 

DFKE is a very standard way of carrying out a key exchange between a pair of parties over a public channel. The privacy of the pair-wise secret keys relies on the hardness of the \textit{Discrete Log Problem} (DLP). Fundamentally, DFKE uses a variation of DLP, known as DHP (Diffie-Hellman Problem). The basic statement of DHP is \textit{"Given $a$, $a^x$, and $a^y$, determine $a^{xy}$"} which has been proved to be computationally hard. It has been successfully used in several other applications such as ElGamal Encryption, Digital Signature, and so on.

To minimize the overhead, in LiPI, the pairwise keys are generated on the fly once in a while and renewed as needed, e.g., when a new node joins or the old keys have been already used a number of times. To reuse the key that is already generated, we use a PRNG seeded by it along with the current iteration sequence number of MiniCast (which is globally the same in all nodes). PRNG ensures a fully random sequence while using the combination of the same key and sequence number and guarantees that an adversary not knowing this seed has only negligible advantage in distinguishing the generator's output sequence from a random sequence.

\textbf{Implementation:} The DFKE process makes use of discrete modular exponentiation. To perform discrete exponentiation of $X^Y \text{mod}~N $ in $log~N$ time, we use (\ref{squaringmodulo}) and appropriately engineer the process, 
e.g., saving and reusing the values using dynamic programming.

\begin{equation}\label{squaringmodulo}
\resizebox{\linewidth}{!} 
{$X^{Y} \bmod N=\left\{\begin{array}{c}\left(X^{\left\lfloor\frac{Y}{2}\right\rfloor} \bmod N\right)^{2} \bmod N \textit{\text {; if } Y \text {is even}} \\ X * \left(\left(X^{\left(\frac{Y}{2}\right)}{\bmod N}\right)^{2} \bmod N\right)\bmod N \textit{\text {; else}}\end{array}\right.$}
\end{equation}

\begin{algorithm}[htbp]
    \small
    \caption{{ \textsc{Algorithmic description of LiPI}}}\label{algo:controller}\vspace{1mm}

    \textbf{\underline{\textsc{Part-1: (Pre-Processing: Offline Phase)}}}\vspace{1mm}
    
    \textbf{\underline{\textsc{Every node $N_i$ in set of nodes $X$}}}\vspace{1mm}
    \begin{algorithmic}[1]

        \State{\textbf{Input:} $P_{ij}$ $\forall$ $N_j$ $\in X \setminus N_i, seq\_no, f_3$} 
        
        \State{\textbf{Output:} $q_{ij}$ $\forall$ $N_j \in X\setminus N_i$} 

        \State{\textbf{Initialise: }    
        $r_{ij}$ $\gets$ 0, 
        $q_{ij}$ $\gets$ 0 }
    
        \vspace{2mm}
    
        \For{$j$ $\gets$ 1 to N $\in$ X $\setminus$ $N_{i}$}         
                \State{$r_{ij} \gets PRNG(P_{ij}||seq\_no)$ $,$} 
                \State{$q_{ij} \gets f_3(r_{ij})$} 
            \EndFor
            
    \vspace{2mm}

    \end{algorithmic}
    
    \textbf{\underline{\textsc{Part-2: (Aggregation: Online Phase)}}}\vspace{1mm}
    
    \textbf{\underline{\textsc{Every node $N_i$ in set of nodes $X$}}}\vspace{1mm}
    
    \begin{algorithmic}[1]
        \State{\textbf{Input:} Secret value $S_i$, Noise $q_{ij}$ $\forall$ $N_j \in X\setminus N_i$, $f_1$, and $f_2$}
         \State{\textbf{Output:}  $S= f(S_1, S_2, ..., S_i, ..., S_N)$}
    
        \vspace{2mm}
        \State{/* \textit{Local computation} */}
               
        \State{Calculate $M_i = f_1(S_i,q_{i1}, q_{i2}, q_{i3}, ..., q_{ik})$}
        
        \vspace{2mm}
        \State{/* \textit{Communication protocol starts} */}
       \State{\textbf{Prepare} a packet with value $M_i$ and place in MiniCast chain}
       \State {Execute all-to-all data-sharing using MiniCast.} 
       \State{/* \textit{Stops after completion of NTX transmission of chain} */}
        \State{Obtains a vector $<M_1, M_2, ..., M_i, ..., M_n>$}
        \State{/* \textit{Communication protocol ends} */}

        \vspace{2mm}
        \State{/* \textit{Local computation} */}
        \State{$S \gets f_2(M_{1}, M_{2}, ..., M_n)$}
    \end{algorithmic}
\end{algorithm}

\begin{algorithm} [htbp]
    \small
    \caption{{ \small \textsc{DF-Key Exchange (DFKE)}}}\label{algo:dfkey}
    
    \textbf{\underline{\textsc{Initiator/Sink}}}\vspace{1mm}
    
    \begin{algorithmic}[1]
  \State{Decide a prime number $p$ and a primitive root modulo of the prime number $g$.}
  
  \State{Initiate a Glossy flood to share $p$ and $g$ with all other nodes.}

    \end{algorithmic}

    \textbf{\underline{\textsc{Every source node $N_i$ in set of nodes $X$}}}\vspace{1mm}
    
    \begin{algorithmic}[1]
        
        \State{\textbf{Input}: Receive $p$ and $g$.}
        
        \State{Decide a secret value $d_{i}$, and compute $V_i = g^{d_{i}}mod~p$}.

        \State{\textbf{Output}: $\forall$ $N_i$, $N_j$ $\in$ $N$ $\exists$ $P_{ij}$}

        \vspace{2mm}
        \State{/* \textit{Communication protocol starts} */}
       
       \State{\textbf{Prepare} a packet with value $V_i$ and place in MiniCast chain}
       \State {Execute all-to-all data-sharing using MiniCast.} 
       \State{/* \textit{Stops after completion of NTX transmission of chain} */}
        \State{Obtains a vector $<V_1, V_2, ..., V_i, ..., V_n>$}
        \State{/* \textit{Communication protocol ends} */}

        \vspace{2mm}
        
        \State{For every other node $N_j$ locally computes the value $((V_{j})^{d_{i}})mod~p$.}
        \State{The common secret keys $P_{ij}$ shared between each pair of nodes, $N_i$ and $N_j$ is computed as $P_{ij} = ((V_{j})^{d_{i}})mod~p = ((V_{i})^{d_{j}})mod~p = (g^{d_{i}.d_{j}})~\text{mod}~p$,}
    \end{algorithmic}
\end{algorithm}
\vspace{-0.8cm}
\subsection{Examples}
Let us consider network-wide summation, i.e., calculation of $S=\sum_{i=1}^{N}S_i$ to be the target aggregation function. Here, both $f_1$ and $f_2$ are summation functions. The noise contribution by a node $N_i$ for node $N_j$, i.e., $q_{ij}$ can be computed directly as $r_{ij}$, if $i<j$, and as $-r_{ji}$ if $i>j$. Thus, $M_i$ is computed using $f_1$ as $S_i + \sum_{i\neq j}q_{ij}$. Hence, $M_i = S_i + \sum_{j=1,i<j}^{n} r_{ij} - \sum_{j=1,i>j}^{n} r_{ji}$. After the MiniCast-based data-sharing, i.e., obtaining all the $M_i$'s from all the nodes, each node can calculate $S$ (i.e., derive $f$) as follows.
\vspace{-0.5cm}
\begin{center}
$$
 \begin{aligned}
\sum_{i=1}^{n} M_{i} &=\sum_{i=1}^{n}\left(S_{i}+\sum_{j=1, i \neq j}^{n} q_{i j}\right) \\
&=\sum_{i=1}^{n} S_{i}+\sum_{i=1}^{n}\left(\sum_{j=1, i<j}^{n} r_{j i}-\sum_{j=1, i>j}^{n} r_{j i}\right) \\
&=\sum_{i=1}^{n} S_{i}
\end{aligned}   
$$
\end{center}

The proposed strategy to achieve PPDA is applicable for a wide variety of aggregation functions. It mainly depends on the availability of suitable $f_1$, $f_2$ and $f_3$. Below we provide some more examples. 

A class of mean functions known as \textit{Quasi-Arithmetic Mean} (QAM) can be readily computed with the help of the proposed strategy. Below we first define QAM and next show the details regarding two possible QAMs, namely, \textit{Arithmetic Mean} and \textit{Geometric Mean}.

\textbf{Quasi-Arithmetic Mean} (QAM): Let $g$ be a function that maps an interval $I$ of the real line to the real numbers and is both continuous and injective. The $\boldsymbol{g}$-mean of $n$ numbers $x_{1}, \ldots, x_{n} \in I$ is defined as $M_{g}\left(x_{1}, \ldots, x_{n}\right)=g^{-1}\left(\frac{g\left(x_{1}\right)+\cdots+g\left(x_{n}\right)}{n}\right)$.  
In order to carry out QAM through LiPI, what is mainly necessary is a relevant transformation function $f_1$ and $f_2$. Let $\vec{x}$ and $\vec{r}$ denote the set of private values and the set of noise values generated by the nodes, respectively. The overall process of privacy-preserving computation of QAM can be expressed through (\ref{eq1}) as follows.

\begin{equation} \label{eq1}
M_{g}(\vec{x})=m^{-1}\left(f_2\left(\frac{1}{n} \sum_{k=1}^{n} f_1\left(g\left(x_{k}\right),f_3\left(\vec{r}\right)\right)\right)\right)
\end{equation}

Here, $g$ needs to be injective in order for the existence of the inverse function $g^{-1}$. Since $g$ is defined over an interval, $\frac{g\left(x_{1}\right)+\cdots+g\left(x_{n}\right)}{n}$ lies within the domain of $g^{-1}$. 

In the following, we discuss two specific examples of QAM. In the definition of the $f_3$ for both of them, we use the function $\textit{Rv}$, which reverses the bits of the binary representation of a string for enhanced security, e.g., $Rv(1101) = (1011)$. 

\textbf{Arithmetic Mean} (AM): AM is one of the QAMs that measures the \textit{central tendency of a distribution}. It proves to be helpful in several statistical analyses. For AM, the function $g$ is defined as \(g({x})=x\). Let \(S_{i}\) be the secret value of a node $N_i$. The joint function to be computed is thus, \(f\left(S_{1},S_{2},\cdots, S_{n}\right)=\frac{\sum_{1=1}^{n} S_{i}}{n}\). With reference to (\ref{eq1}), AM can be calculated using LiPI using the following definition of $f_2$, $f_2$ and $f_3$.

\begin{center}
    $f_1(g(x_{k}),f_3(\vec{r})) = f_1(S_{i},f_3(\vec{r})) = M_{i} =  S_{i}+\sum_{1=1}^{n} q_{ij}$\\
    $f_{2}(M_{1},M_{2},\cdots, M_{n}) = \sum_{1=1}^{n} M_{i} = \sum_{1=1}^{n} S_{i}$
\end{center}
\resizebox{\linewidth}{!}
{$q_{ij}=f_{3}(r_{ij})= \begin{cases}
-\textit{PRNG}(P_{ij}||Seq\_no.)-\textit{PRNG}(\textit{Rv}(P_{ij})||Seq\_no.)    & i < j\\
\textit{PRNG}(P_{ij}||Seq\_no.)+\textit{PRNG}(\textit{Rv}(P_{ij})||Seq\_no.) & i > j
\end{cases}
$}
\\

\textbf{Geometric Mean} (GM): GM also measures the central tendency, but it is more suited for when the data values are multiples or exponential to each other in nature. In such a case, the arithmetic mean fails to give the actual central tendency. GM of a set of data $\left\{a_{1}, a_{2}, \ldots, a_{n}\right\}$ is computed as follows:
\begin{equation} \label{eq2}
\left(\prod_{i=1}^{n} a_{i}\right)^{\frac{1}{n}}=\sqrt[n]{a_{1} a_{2} \cdots a_{n}}
\end{equation}

With a similar line of arguments, the three functions for the computation of GM in a privacy-preserving manner using LiPI are defined as follows.
\begin{center}
$f_1(S_{i},f_3(\vec{r})) = M_{i} =  S_{i}\times \prod_{1=1}^{n} q_{ij}$\\
$f_{2}(M_{1},M_{2},\cdots, M_{n}) = \prod_{1=1}^{n} M_{i}=\prod_{1=1}^{n} S_{i}$
\end{center}

\resizebox{\linewidth}{!}
{$q_{ij}=f_{3}(r_{ij})=  \begin{cases}\textit{PRNG}(P_{ij}||Seq\_no.)^{-1}  \times \textit{PRNG}(\textit{Rv}(P_{ij})||Seq\_no.)^{-1}  & i < j\\
\textit{PRNG}(P_{ij}||Seq\_no.) \times \textit{PRNG}(\textit{Rv}(P_{ij})||Seq\_no.)& i > j
\end{cases}
$}
\\

Note that calculation of GM can also be done by first applying a logarithmic transformation on the private value $x_i$, (i.e., \(g(x_{i})=log(x_{i})\)) and later applying the functions used for calculation of AM. A variety of QAM, including \textit{harmonic mean, power mean, log semi-ring}, etc., can also be calculated in a privacy-preserving manner through LiPI. Apart from that, it can be also used for other different statistical measures too provided proper $f_1$, $f_2$ and $f_3$ are first derived.

\section{Design of the System}
\label{sec:failure}

This section describes the design of a full system that binds all the components together. The basic process is fundamentally quite simple. It starts with a synchronization round by Glossy followed by local computation of the $M_i$'s. Subsequently, a round of MiniCast is executed for sharing the $M_i$'s with each other. However, the masking function used in LiPI in each node incorporates the noise contribution for all the other legitimate nodes in the system. Thus, for successful de-masking, participation of all the source nodes is highly essential. Any failure or non-participation (intentional/unintentional), thus, would cause incorrect/incomplete computation of the aggregation. The ability to handle such node failure is one of the prime issues in any practical system in general. It is a challenging task in a decentralized system and is not addressed satisfactorily in most of the existing works \cite{LPDA, LVPDA, 4215819, 6208686, PPMP}. In LiPI, we exploit the specialty of both ST and the proposed aggregation mechanism to support node failures up to a certain extent. Below we explain the recovery mechanism with two possible scenarios.

\textbf{Partial data-sharing}: In a specific round of MiniCast (aggregation), for some reason, after initial participation in the data-sharing process, a legitimate source node (say $x$) may fail and become unable to complete the data-sharing process (i.e., fail to complete the required number of transmissions of the complete chain in MiniCast). However, with a high chance, the data would be received by at least one \textit{healthy node} (say $y$) through the initial transmissions from $x$. Later the data of $x$ will be shared through $y$ which will enable the other nodes to correctly compute the final aggregated value. Proper exploitation of redundant and broadcast-based transmissions thus enables an ST-based system to preserve the values shared by a node even if it fails after at least one successful transmission.

\textbf{Non-participation}: A valid source node may completely seize to participate in the data sharing process after participating in DFKE. However, due to the transparency regarding the ownership of each of the sub-slots in MiniCast, every node would be able to quickly detect how many other nodes and explicitly which node did not share their data. We augment the protocol LiPI with a systematic recovery procedure to mitigate this situation. The initiator, in such cases, disseminates a final list of nodes that could not participate in the last round. Based on this, every node recomputes the value of $M_i$ (see Algo. \ref{algo:controller}, Part-2, line 4) considering the noise values ($q_{ij}$) for only those nodes whose ids are not present in the shared list. Followed by this a recovery MiniCast round is executed through which the new $M_i$'s are shared and the nodes correctly compute the final result.

\subsection{The Process in Action}
\label{sec:protocolrun}

The timing diagram and the flow diagram explaining the overall process are provided in Fig. \ref{fig:state} (Part (a-d)). The overall process runs in a periodic fashion \cite{sync_survey}. The pair-wise key exchange process DFKE is executed only at a pre-defined interval or in special situations, e.g., joining/leaving of a node. PPDA process can start only after at least one round of DFKE has been successfully executed (i.e., the $P_{ij}$ values are set). To cover all the possible failure cases as depicted in the previous section, at most, two back-to-back full MiniCast instances are executed with a Glossy \cite{glossy} phase at the beginning of both. The first Glossy phase is used for overall network synchronization and dissemination of control data, e.g., a list of valid source nodes, etc., while the second Glossy phase is used for disseminating the list of nodes that are missing in the former data-sharing step, so that in the subsequent MiniCast round the remaining nodes can share the recomputed $M_{i}$'s. Note that the second instance of Glossy and the MiniCast for recovery are needed only when the nodes/initiator detects the absence of some node(s) in the first round.

\textbf{Implicit advantage of ST over AT: } Note that ST-based communication mechanism is the backbone of LiPI, bringing forth certain special advantages apart from low latency and low energy consumption. Under traditional AT-based communication, because of comparatively higher delay, the strictly time-bounded network-wide end-to-end operation is hard to implement. Unpredictable delays in communication may result in a serious breach of privacy. Consider a node $x$ which failed to successfully communicate its $M_x$ to the other nodes in round $r$ and it resulted in the execution of a recovery procedure in round $r$. However, due to network delay, $M_x$ may arrive even after completion of the recovery procedure in round $r$, which can be exploited by any node to reveal the private data of $x$. Strictly time-bounded and time-separated network operations in ST completely rule out such possibilities.

\section{Evaluation}
\label{sec:evaluation}
\textbf{Experimental-Setup}: We implement LiPI in the Contiki Operating System for TelosB devices and extensively experiment in both Contiki Network Simulator \textit{Cooja}~\cite{cooja} as well as publicly available IoT/WSN testbeds FlockLab~\cite{flocklab} and DCube\cite{dcube}, which contains 24 and 47 TelosB devices, respectively. DCube testbed has two loosely connected islands containing 31 and 16 nodes. To isolate the possible reliability issues related to the execution of network protocols in all our experiments, we mostly use the larger island, having 31 nodes.

\subsection{Parameters and Metric}
The parameter NTX in MiniCast determines how many times a node forwards or transmits the chain of packets. In ST-based protocols, NTX determines mainly the reliability. However, in MiniCast, since it starts the floods from different source nodes, a less reliable operation implies a limited outreach of each individual flood. NSSS exploits this to optimize its performance. In all other protocols, we need reliable all-to-all data-sharing and hence set the NTX values accordingly (i.e., the minimum NTX needed for a specific network setting to achieve global outreach for every flood).

The metrics used to measure the efficiency and robustness of LiPI and other strategies are described below.

\begin{itemize}

\item \textbf{Latency:} It is the time taken for a node to obtain the final aggregation value. It considers both the computation and communication phases of the protocol. 

\item \textbf{Radio-on time:} It is the total time for which a node keeps its radio ON for the data sharing operation. Radio-on time depends on the value of the NTX. In resource-constrained IoT devices, Radio-on time directly reflects the energy consumption as radio is supposed to be the most power-hungry component in the device. 
\end{itemize}

\begin{figure}[!htbp]
    \centering  
   
    \includegraphics[width=\linewidth]{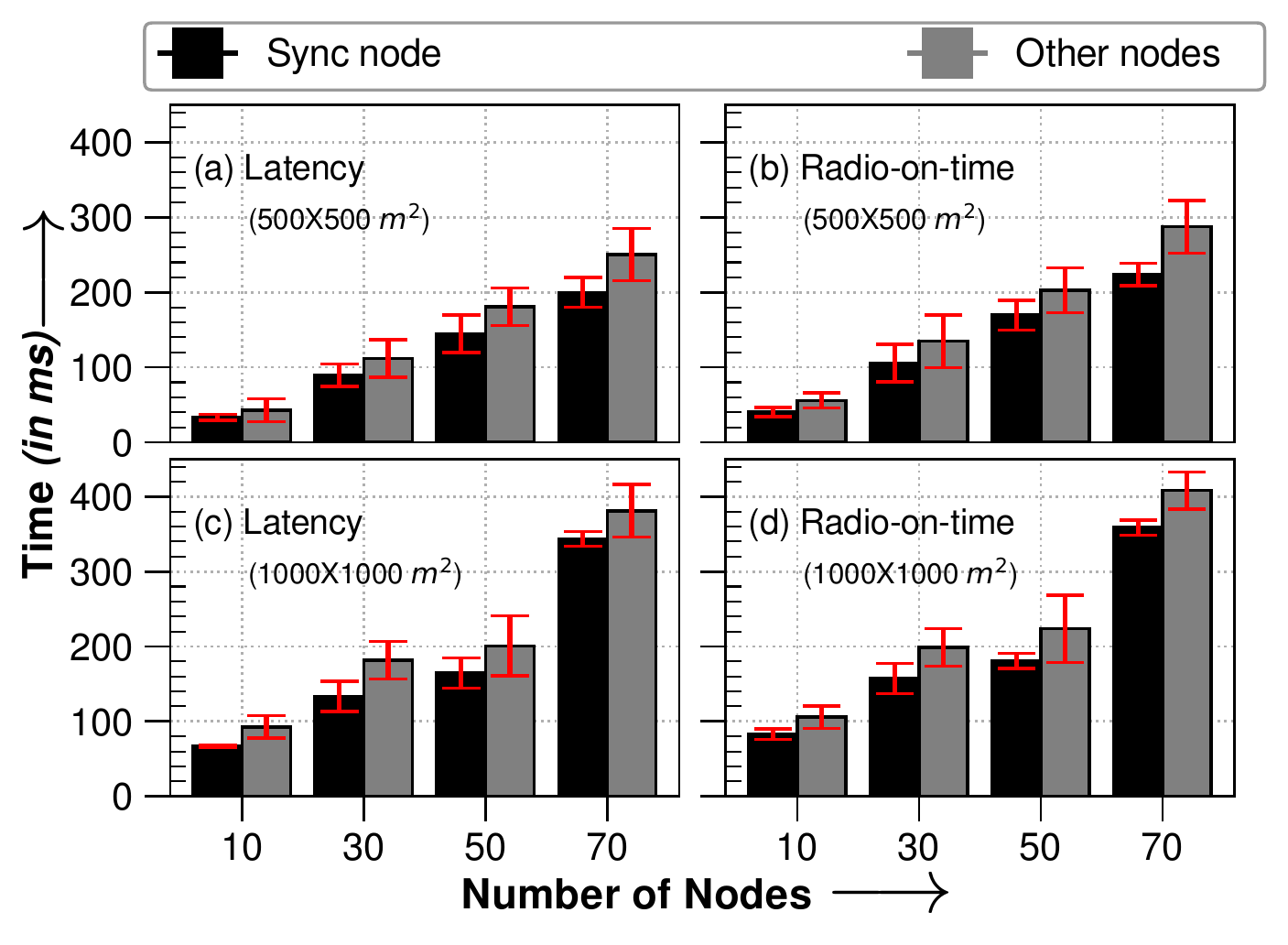}
    \vspace{-0.8cm}
    \caption{Latency (Part (a,c)) and Radio-on time (Part (b,d))  in executing LiPI-based calculation of sum values in IoT/WSN setting comprised of the different number of nodes and spread over different areas in Cooja.}
    \label{no_failures_siml}
\end{figure}

\begin{figure}[!htbp]

\includegraphics[width=\linewidth]{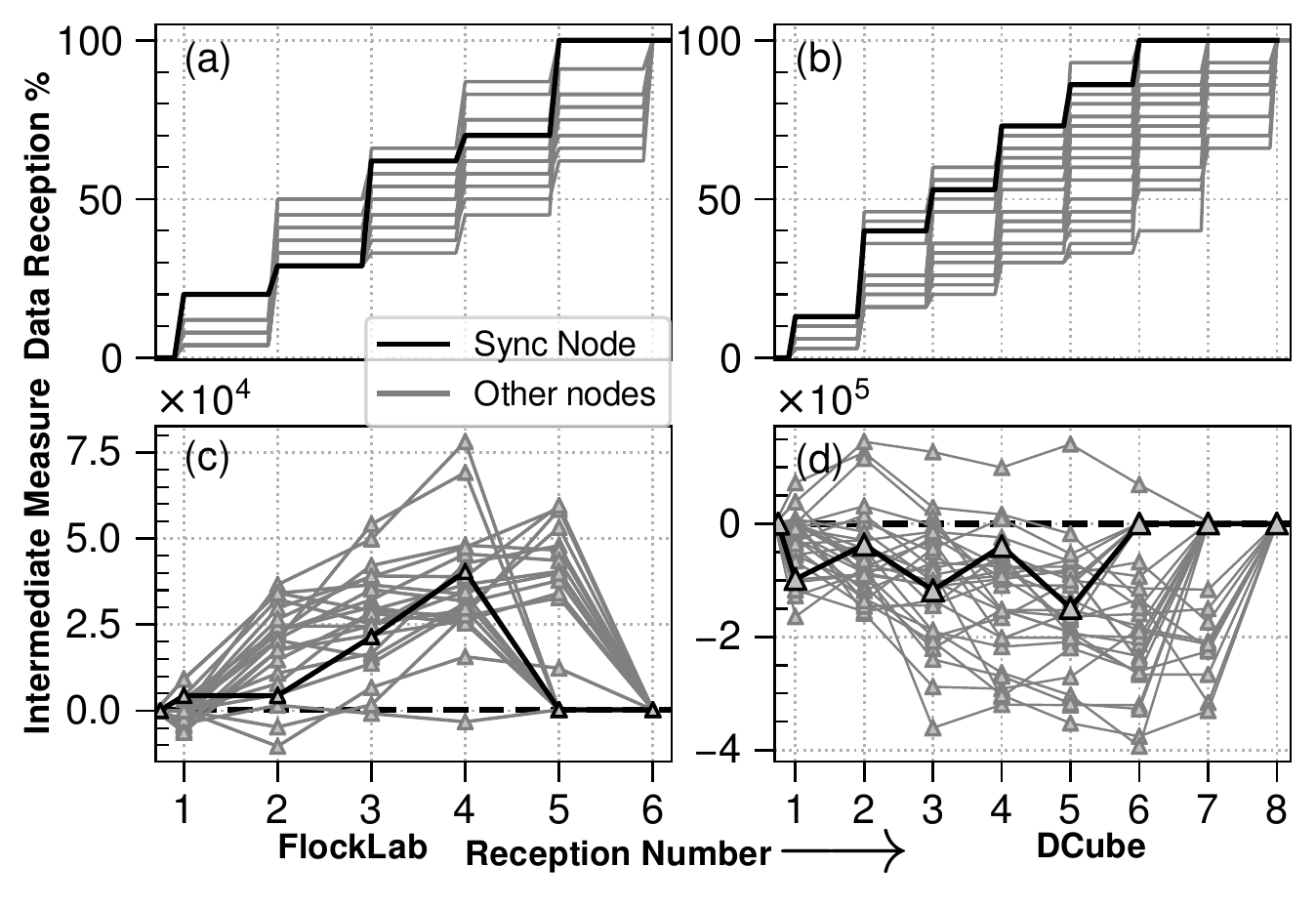}
\vspace{-0.8cm}
\caption[Fig.]{Percentage of data received and the intermediate aggregated value after each chain reception}
\label{randomness}
\end{figure}

\subsection{Basic study}

We test the strategies first with Cooja. Each experiment is executed for at least 1000 iterations, and the metrics are computed as an average over all the iterations and all the nodes. The error bars in the results reflect the standard deviation. For simplicity in all the evaluation experiments, we assume the target aggregation function to be the summation of all the secret values. In any ST-based protocol, an initiator (here referred to as \textit{sync}) node starts the whole process. In general, any node can work as an initiator, and a new initiator can get elected when the existing one fails. However, because of its special role, in all the results, we separately report the Latency and Radio-on time of the initiator/sync.

\textbf{Effect of the number of nodes:} LiPI is first executed in Cooja with 10, 30, 50, and 70 no. of nodes randomly spread over a different area. Fig.~\ref{no_failures_siml}(a) and \ref{no_failures_siml}(c) show the Latency in both sync and other nodes for two different areas \( 500\times500\), and \(1000\times1000\) (all in $m^{2}$) respectively. Radio-on-time values are shown in \ref{no_failures_siml}(b) and \ref{no_failures_siml}(d). Almost a linear increase in the metrics with a number of participants is visible in both cases. However, with the increase in the deployment area, the diameter of the network increases, which makes the underlying data-sharing process put more effort into executing all-to-all data-sharing. This is reflected in the considerable rise in the Latency and Radio-on time for the 70-node network in 1000x1000 sq. meters compared to the other settings.

\textbf{Evolution of the aggregation}: 
In this experiment, we minutely check how the aggregation values evolve with time in a single round in each node and whether the computation of the partial results in the nodes leaves any chance for the private values to get revealed. Fig.~\ref{randomness}(a) and \ref{randomness}(b) show the percentage of the data received (\textit{Data Reception \%}) in some of the selected nodes at regular intervals during the execution of two sample instances of LiPI in two different testbeds, FlockLab, and DCube. Fig.~\ref{randomness}(c) and \ref{randomness}(d) show the corresponding intermediate calculation of the aggregation in those nodes. The behavior of the sync node is shown in a thick black line. The target sum value is shown in thick dashed black lines in Fig.~\ref{randomness}(c) and \ref{randomness}(d). Intermediate values, i.e., the partially computed aggregation values, make a random transition over a wide range of values, and hence the final value, although they are positive numbers, they appear to be a very small value (and appear to be coinciding with zero.). In both cases, we assume that each node shares its node-id as the secret value. Thus, in Fig.~\ref{randomness}(c) and \ref{randomness}(d), the final sum converges to a target value that is the sum of the node-ids, i.e., $\frac{n(n+1)}{2}$ (N being the total number of nodes). In particular, in FlockLab and DCube, the sum obtained is 300, and 496 (sum of the virtual ids from 1 to 24 and 1 to 31, respectively). In all these experiments, random transitions of the intermediate values ensure orthogonality of the same w.r.t. the finally derived aggregation value.

\subsection{Comparison of LiPI with other PPDA strategies}
\label{subsec:compareimplementation}
In order to compare the performance of LiPI, we select three different state-of-the-art PPDA strategies, namely, PPMP~\cite{PPMP}, SSS~\cite{iccsss} and NSSS~\cite{goyal2022multiparty}. However, none of the works provide any implementation for low-power hardware devices. In addition, there is no attempt so far to incorporate ST based framework to realize any PPDA strategy. Therefore, to make a fair comparison, we implement all three strategies in the same ST-based framework that we use for LiPI. In the following, we describe them in detail.

\subsubsection{PPMP}
PPMP satisfies a significant fraction of the properties that are expected from a PPDA strategy and are targeted in this work, too (see Table~\ref{fig:workscomparison} in Section \ref{sec:related}). It assumes a special circular arrangement of the nodes where every node $i$ is assigned with exactly two adjacent nodes (referred to as $(i+1)$ and $(i-1)$). The whole task is carried out in two rounds of communication/data-sharing, referred to as the \textit{key exchange} and the \textit{aggregation}. In key-exchange round every node $N_i$ first picks up a random number $r_i$ and computes $g^{r_{i}}$ ($g$ being a globally known group generator) which is shared with its right and left adjacent nodes, $N_{i+1}$ and $N_{i-1}$, respectively. Later, node $N_i$ obfuscates it's secret value $x_i$ and obtains $C_i$ using a function $f$ involving $x_i$, $p$ and $R_i$ (e.g., for calculation of sum as from the final aggregation $C_i = f(x_i, p, R_i) = (1+x_i.p).R_i$), where $p$ is a globally decided prime number and $R_i$ is the ratio of the two values received by $N_i$ from its right and left adjacent nodes, i.e., $(g^{r_{i+1}}/g^{r_{i-1}})^{r_{i}}$. In the aggregation round, every node shares $C_i$ with the entire network. Followed by this, the nodes take the product of the $C_i$-s received from all other nodes and apply principles of modular arithmetic (modulo $p^2$) and polynomial expansions to obtain the final result \cite{PPMP}.

We implement PPMP using MiniCast. In the key-exchange round, PPMP needs to maintain a circular arrangement where every node is supposed to share the keys with its two adjacent nodes (left and right). Under a wireless decentralized \textit{ad hoc} setting, we realize this with the help of node-id, e.g., a node having id $i$ is assumed to have the nodes with ids $(i+1)\%n$ and $(i-1)\%n$ as the right and left adjacent nodes, respectively, irrespective of their physical location. Therefore, the key-exchange round is executed through a MiniCast-based all-to-all data-sharing. Subsequently, the aggregation is executed through another complete round of MiniCast. Note that DFKE in LiPI can be compared with the key-exchange round in PPMP. Although both need the same communication effort, in LiPI, it is enough to execute DFKE once after several rounds of aggregation. However, in PPMP, the fresh key exchange has to precede every aggregation round.

Note that PPMP cannot use PRNG even (as LiPI does) to reuse the same key for multiple rounds for obvious reasons. In case the same key is repeated in consecutive rounds of aggregation in PPMP, a \textit{curious} adversary would become able to reveal the temporal evolution of the private values for any target node. For instance, the ratio of the obfuscated data by a node $N_i$ at time-step $t$ and $t+1$ can be calculated as $\frac{C(t)}{C(t+1)}=\frac{(1+x(t)p)(g^{r_{i+1}}/g^{r_{i-1}})^{r_{i}}}{(1+x(t+1)p)(g^{r_{i+1}}/g^{r_{i-1}})^{r_{i}}}$, which can be simplified as $\frac{C(t+1)}{C(t)}=\frac{(1+x(t+1)p)}{(1+x(t)p)}$ when the nodes repeat using the same $r_i$'s. This can be easily exploited by adversaries to reveal the temporal properties of private values, causing a breach of privacy. In contrast, in LiPI, the same pairwise keys can be repeatedly used for several rounds as the obfuscation procedure is much stronger and depends on the collaborative contributions from all the members making the results completely random and no possibility of inferring temporal evolution.

\begin{table}[t]\centering
\caption{Comparison summary}
\begin{tabularx}{\linewidth}{ | c | c | c | c |X | }
\hline \hline \textbf{Algo.} & \textbf{Encryption } & \textbf{Comm. } & \textbf{Share gen.}& \textbf{Collusion }\\
Type & Cost & rounds & Cost & threshold\\

\hline SSS\cite{iccsss} & Pair-wise & $2$ & $\Theta(n^{2})$ & $n-1$ \\
\hline NSSS\cite{goyal2022multiparty} & Pair-wise  & $2$ & $\Theta(dn)$ & $d-1$ \\
\hline PPMP\cite{PPMP} & Open & $2$ & $O(M(n)k)$ & $1$ \\
\hline LiPI & Open & $1$ & $\Theta(n)$ & $n-2$ \\
\hline \hline
\end{tabularx}
\label{comparison}
\end{table}

\subsubsection{Shamir's Secret Sharing (SSS)}
SSS has been used in many works for SMPC, which can be directly used for PPDA. In SSS, every node first decides a local n-degree polynomial $P_i(x)$ where $n$ is the number of nodes in the system. The secret key $S_i$ of $N_i$ is considered to be the constant terms in $P_i$, i.e., $P_i(0)$. $P_i$ is first evaluated at $n$ distinct predefined points ($y_1, y_2, y_3,...,y_n$) and the values are next shared with the nodes in the network using secure channel (i.e., $P_i(y_j)$ is shared by $N_i$ only with $N_j$ through a secured channel). To maintain uniformity, we implement this using MiniCast, where each node puts data in ($n-1$)-distinct sub-slots encrypted (AES-128) using pre-agreed pair-wise secret keys (we use DFKE for settling the pairwise keys). After completion of MiniCast, every node $N_i$ decrypts all the $n$-1 values received and sum up locally to derive $K_i=\sum_{i=1}^{n}(P_i(y_j))$. Next, the second round of MiniCast (\textit{reconstruction-round}) is used to share the values of $K_i$ from each node with each other in plain text. This finally constructs a polynomial $\mathcal{P}$ using lagrange-interpolation in every node which represents the sum of all the polynomials adopted by all the participating nodes. Thus the constant term in $\mathcal{P}$ happens to be the sum of the secret values i.e., $\sum_{i=1}^{n}(S_i)$.\\

\subsubsection{Neighborhood-based Shamir's Secret Sharing (NSSS)}
Implementation of SSS, even through our proposed ST-based strategy, is extremely communication intensive. In particular, the size of the individual chain used in SSS is $O(n^2)$ as it uses all-to-all communication for both data-sharing and reconstruction-round. The work \cite{goyal2022multiparty} demonstrates an optimized version of SSS where a node instead of interacting with everyone else in the network, only involves its neighbors. Specifically, it assumes a $d$-degree polynomial of degree lesser than n, i.e., the total number of nodes, in such a way that in the data-sharing round, it is sufficient for a node to share its data only with its neighbors. To accomplish this, the parameter NTX in MiniCast is appropriately adjusted to tame the outreach of the propagation of the data from each node only up to a few hops. This substantially reduces the communication overhead compared to SSS. We refer to this version as \textit{Neighborhood based SSS} (NSSS). However, it still requires encryption of packets before transmission in the sharing phase of MiniCast. Moreover, the re-construction phase also requires an all-to-all interaction among the participants.

The summary of the minute comparison of LiPI w.r.t. PPMP, SSS, and NSSS is provided in Table~\ref{comparison} and explained below.

\textbf{Encryption:} Both SSS and NSSS require explicit encryption of data to implement a secure channel between pairs of nodes. In contrast, as already described, LiPI and PPMP do not need any encryption.

\textbf{Collusion:} SSS and LiPI have the highest collusion resilience (collusion-threshold is $n$-1, and $n$-2) as both mandate the participation of almost all the nodes in the computation of the aggregation. NSSS improves over SSS by restricting the communication only among the neighbors. Although it makes the protocol simpler, \textit{collusion threshold} is decreased.
PPMP exhibits the least resilience against collusion. Because of the circular arrangement of the nodes in PPMP, a node shares data with two other nodes. Let us consider every node $N_i$ shares $C_i=(1+x_{i}*p)*(g^{r_{i+1}}/g^{r_{i-1}})^{r_{i}}$ (where $x(i)$ is the private data) with its two adjacent nodes. Now, the adjacent nodes can collude with each other to calculate $(g^{r_{i}})^{r_{i+1}}/(g^{r_{i}})^{r_{i-1}}$ as $r_{i+1}$ and $r_{i-1}$ are private to them and $g^{r_{i}}$ is already received in the key-exchange round. Thus, in PPMP, the collusion threshold is just 1.

\textbf{Data-sharing complexity:} Finally, LiPI substantially minimizes the communication requirement. Ideally, it just needs a single round of MiniCast-based all-to-all data-sharing. However, in all the other strategies, even when there is no node failure, it requires multiple rounds of MiniCast. Moreover, due to the use of restricted data-sharing, a node needs to reserve multiple sub-slots in MiniCast, which substantially increases the cost of communication.

\begin{figure}[t]
    \centering
    \includegraphics[width=\linewidth]{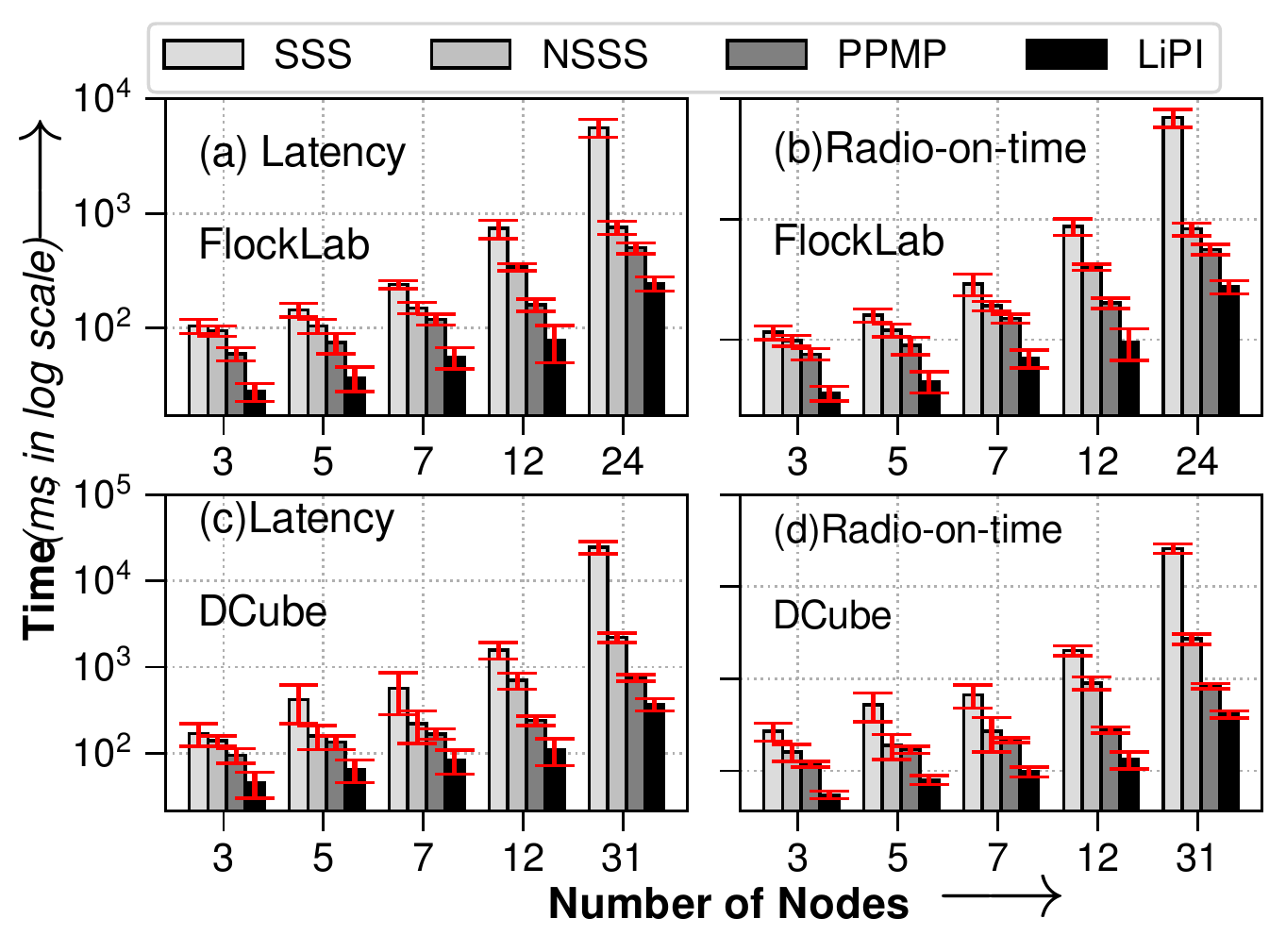}
    \vspace{-0.8cm}
    \caption{Execution of SSS, NSSS, PPMP, and LiPI in FlockLab and DCube, respectively}
    \label{compare}
\end{figure}

\textbf{Validation in Testbed:} We validate the simulation and theoretical findings through extensive studies with the two IoT/WSN testbeds. All the four PPDA strategies are executed to obtain the sum of the randomly assumed private values by the nodes. A comparison of the performance of the protocols in terms of Latency and Radio-on time is depicted in Fig.~\ref{compare}. It can be observed that LiPI outperforms all the other strategies in both metrics. In FlockLab, on average, LiPI achieves PPDA up to 51.7\%, 68.56\%, and 81.91\% faster and consumes 50.49\%,  66.49\%, and 80.35\% lesser energy compared to PPMP, NSSS, and SSS, respectively. Similarly, in DCube, it accomplishes the task up to 52.04\%, 71.41\%, and 87.3\% faster and consumes 51.72\%,  70.89\%, and 86.7\% lesser energy compared to PPMP, NSSS, and SSS, respectively.

\begin{figure}
    \centering
        \includegraphics[width=\linewidth]{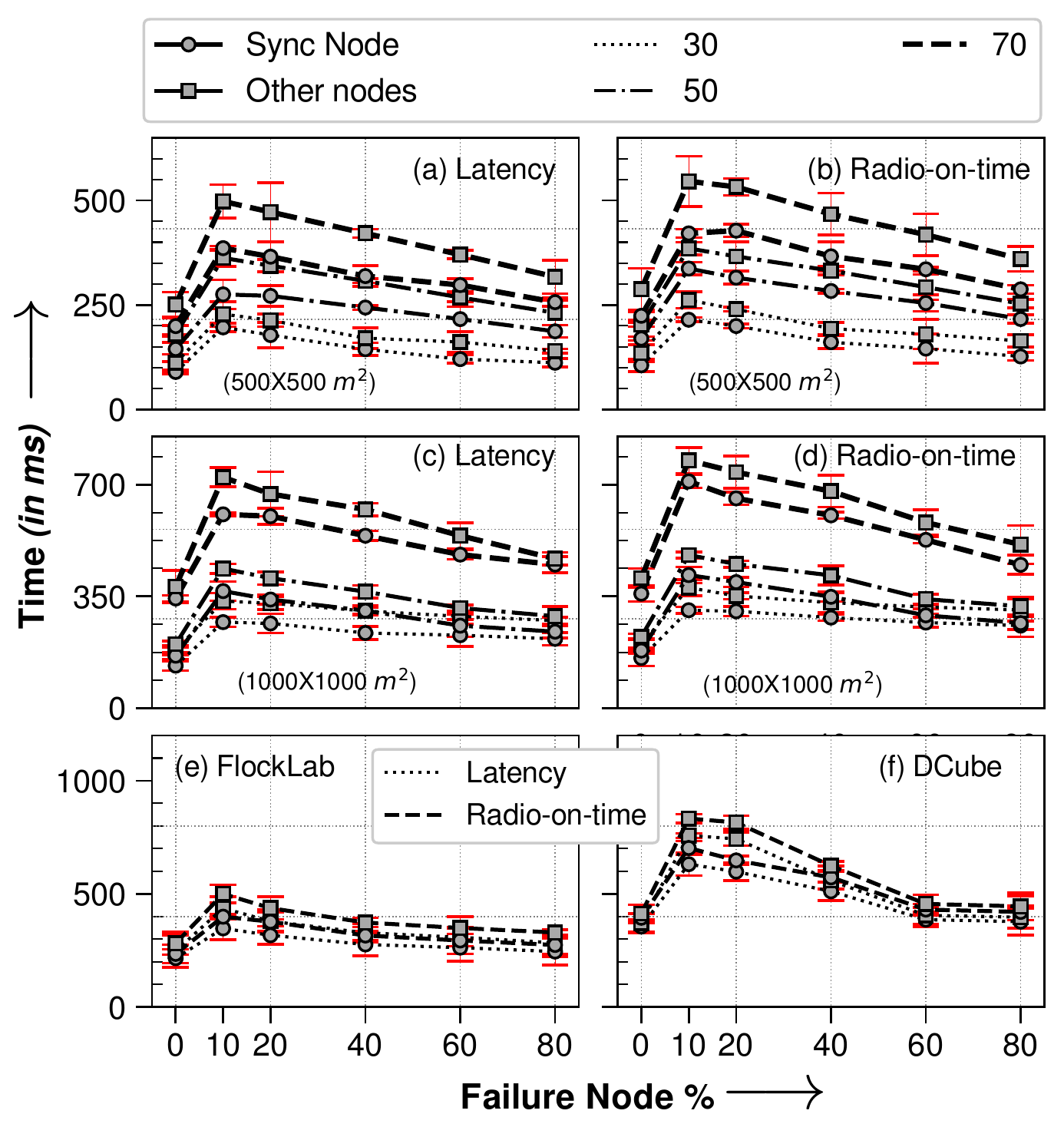}
        \vspace{-0.8cm}
    \caption{Latency and Radio-on time in execution of LiPI for different fractions of node failures.}
        \label{failures}
\end{figure}

\subsection{Node-failures} 
Node-failure is rarely studied in the existing PPDA works, as shown in Table~\ref{fig:workscomparison}. Initially, we analyze the performance of LiPI in various different Cooja configurations when multiple valid nodes fail to participate in the data-aggregation process after successfully exchanging keys in DFKE. Any node failure is mitigated through subsequent correction or re-computation of the aggregation. This naturally increases the communication cost. However, since it has already been shown that communication cost in LiPI is substantially lesser than the other strategies, the impact of node failures is also naturally much lesser. Fig.~\ref{failures}(a-d) shows the results in two Cooja configurations where different numbers of nodes (30, 50, and 70) are spread over 500 x 500 and 1000 x 1000 sq. meters. Even for a single node failure, since the data-sharing process needs to run once again to do the necessary correction, the time and energy consumption straightaway doubles which is visible in all the results. However, the interesting point to note here is that as more nodes start failing, the chain length, i.e., slot-length in the MiniCast, drops which results in lesser Latency and Radio-on time as long as the network remains connected and does not cause any massive increase in the network diameter. 

The same study is also carried out in testbeds FlockLab and DCube where a similar scenario is visible, i.e., the Latency and the Radio-on time substantially rise up for even a smaller percentage of node failure while drop-down slowly as more nodes start failing. Results are shown in Fig.~\ref{failures}(e) and Fig.~\ref{failures}(f), respectively. In all our experiments, we consider a random node failure model over the whole network uniformly, which mostly results in a connected network and does not increase the diameter of the network noticeably.

In the following, we discuss the possible ways node failure can be handled in the other PPDA works. The work PPMP~\cite{PPMP} does not provide any fault-tolerance strategy to mitigate node failures/dropouts. The way we implement PPMP with an ST-based framework can easily detect node failure, as already discussed. However, despite this detection, the nodes will have to rearrange themselves in a circular setting to  continue the execution of the protocol. This would involve re-assignment of virtual-ids and setting the right and left adjacent nodes implicitly or explicitly. 

In contrast to this, both SSS and NSSS support fault tolerance. If a node fails before the data-sharing round, it does not affect the process. If it happens after the data-sharing and before reconstruction, then it depends on the degree of the polynomial used. If a sufficient number of nodes are there to reconstruct, then there is no need to repeat the data-sharing round. This is satisfied in case the degree of the polynomial is lesser than the number of nodes present (after node failure). However, if the number of available nodes is lesser than the degree, then both the sharing and re-construction rounds need to be repeated completely.

\section{Conclusion}
\label{sec:conclusion}
Aggregation of data is a very common operation in any decentralized smart system in general. However, the common model of obtaining aggregated value through sharing the data from each node to a centralized node or sharing data with each other incurs the possibility of breach of privacy. Existing solutions in order to resolve such issues mostly adopt various means involving computation intensive homomorphic-encryption, complex data-obfuscation, or even communication-intensive multi-party computation. These solutions are too heavy to be useful or even applicable for the low-power devices that compose a significant part of an IoT-assisted smart system. In this work, we propose a simple, lightweight protocol LiPI for privacy-preserving-data-aggregation through collaborative data-obfuscation. In addition, we also propose a Synchronous-Transmission based framework where we implement LiPI as well as three other strategies to achieve the same goal. Through extensive simulation and testbed-based experiments, we demonstrate that LiPI outperforms the existing state-of-the-art strategies by a wide large margin.

The collaborative data-obfuscation process used in LiPI uses the contribution from every node in the network, which mandates the use of an all-to-all data-sharing. However, this can be relaxed by allowing only the neighbors of a node to contribute to obfuscating the private data of a node. Although such a setting would reduce the communication cost, it would also pull down the collusion-resilience of the protocol. In addition, the traditional divide-conquer-and-merge strategy can also be applied with LiPI to make it scalable. We consider these works as part of our immediate future steps in this direction.

\bibliographystyle{IEEEtran}
\bibliography{ref}

\IEEEpeerreviewmaketitle

\end{document}